\documentclass[12pt]{iopart}
\pdfoutput=1

\usepackage{graphicx}
\usepackage{upgreek}
\usepackage{textcomp}

\begin{document}

\title{Tunable cavity coupling of the zero phonon line of a nitrogen-vacancy defect in diamond.}

\author{S. Johnson$^1$, P. R. Dolan$^1$, T. Grange$^2$, A. A. P. Trichet$^1$, G. Hornecker$^2$, Y. C. Chen$^1$, L. Weng$^1$, G. M. Hughes$^1$, A. Auff\`{e}ves$^2$ and J. M. Smith$^1$}

\address{$^1$ Department of Materials, University of Oxford, Parks Road, Oxford OX1 3PH, UK}
\address{$^2$ Universit\'e Grenoble-Alpes \& CNRS, Institut N\'{e}el, 38000 Grenoble, France}
\ead{jason.smith@materials.ox.ac.uk}

\begin{abstract}
 We demonstrate the tunable enhancement of the zero phonon line of a single nitrogen-vacancy color center in diamond at cryogenic temperature. An open cavity fabricated using focused ion beam milling provides mode volumes as small as 1.24 $\mu$m$^3$. In-situ tuning of the cavity resonance is achieved with piezoelectric actuators. At optimal coupling to a TEM$_{00}$ cavity mode the signal from individual zero phonon line transitions is enhanced by about a factor of 10 and the overall emission rate of the NV$^-$ center is increased by 40\% compared with that measured from the same center in the absence of cavity field confinement. This result is important for the realization of  efficient spin-photon interfaces and scalable quantum computing using optically addressable solid state spin qubits.
\end{abstract}

\section{Introduction}
Coupling of fluorescence from nanoscale quantum systems to optical microcavities provides a means to control the emission process and can be an essential element of nanophotonic device applications. The negatively charged nitrogen-vacancy (NV$^-$) defect in diamond is an example of a solid state system that has gained significant attention in recent years as a quantum spin register \cite{Dutt07,Childress08} and nanoscale sensor \cite{Balasub08,Rondin14,Toyli13} due to its long spin coherence times and capacity for optical manipulation and readout. The recent demonstration of quantum error correction in an NV$^-$ defect \cite{Waldherr14} provides a sound basis for using these systems in practical quantum information technologies. However a coherent spin-photon interface, necessary for many quantum technologies, can not be achieved if information is leaked via interaction with phonons, and so in the NV$^-$ center only the `zero phonon line' (ZPL) transition at 1.945 eV (637 nm) may be used. The low Debye Waller factor (DW $\simeq0.04$) of the center imposes a significant limitation, since most spontaneously emitted photons are useless. The enhancement of this transition and its efficient coupling to external optics are therefore important challenges to which microcavities are well suited. In particular is an essential requirement for generating large scale entangled states between spatially separated defects connected via photonic networks, for which proof of principle experiments have been achieved \cite{Bernien13} but further development is hampered by low entanglement efficiency. Success in this endeavor will provide a route to the realization of scalable quantum computers based on optical networks of electronic and nuclear spins. \cite{Raussendorf01,Benjamin09}
\begin{figure}[b]	
	\begin{center}
		\includegraphics[width=14cm]{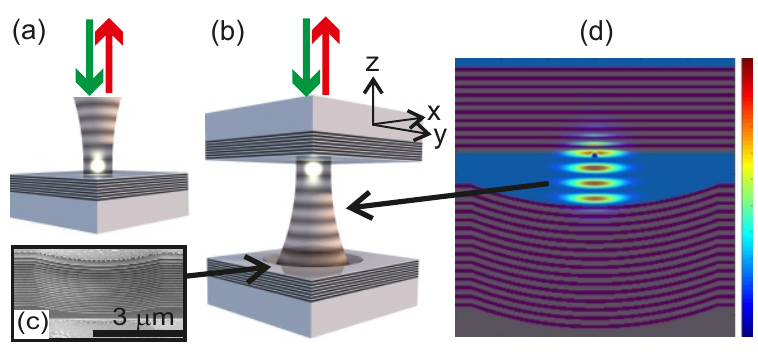} %depending on the latex compiler, you can omit the file extension
		\caption{Experimental configurations for characterisation of the same NV center in (a)  the `out-of-cavity' and (b) the `in-cavity' measurements. (c) Scanning electron micrograph of a cross section through the Bragg coating of the concave mirror. (d) False color-scale plot of the electric field intensity distribution through a cross-section of the $q$ = 4 TEM$_{00}$ cavity mode calculated using FDTD modelling. The refractive index profile is included for clarity, with grey (purple) regions corresponding to the low (high) index layers of the Bragg mirrors.}
		\label{fig:example}
	\end{center}
\end{figure}

Resonant coupling of the NV$^-$ ZPL to micro-ring resonators \cite{Faraon11} and photonic crystal cavities \cite{Faraon12,Li15} has been demonstrated to provide effective enhancements, but the monolithic structure of these cavities prevents positioning of the emitter at the heart of the cavity mode in-situ. Tuning of the cavity mode to the ZPL resonance can be achieved in these systems by progressive condensation of gas molecules onto the structure, thus increasing the mode volume and red-shifting the resonance, but it is difficult to optimise tuning using this procedure and it has limited scope for use in device applications. Open cavities \cite{Trupke05,Steinmetz06,Dolan10, Hunger10} provide a flexible approach to cavity-coupled devices that permit full in-situ alignment and tuning, combined with efficient coupling to external optics. Here we demonstrate control over the emission from a single NV$^-$ center by coupling its ZPL to the resonant mode of an open microcavity. The open cavity geometry allows direct comparison between the emission properties of the same defect in and out of the cavity, thus providing unambiguous evidence of the effect of the cavity mode. We demonstrate clear tunable enhancement of the ZPL emission and reduction of the fluorescence lifetime of the defect in a controlled manner, and analyze our results in terms of the Purcell effect acting on the ZPL and the phonon sideband (PSB). Our work builds on recent demonstrations of room temperature coupling of NV$^-$ defects to open cavities \cite{Albrecht13, Albrecht14, Kaupp13}.

\section{Experimental Method}

The open microcavities are of a plano-concave design, the concave features produced by focused ion beam milling of a fused silica substrate \cite{Dolan10}. These cavities support a Hermite-Gauss mode structure with TEM$_{00}$ modes that can be effectively mode-matched to an external Gaussian beam. The radius of curvature of the concave mirror used here is 7.6 $\mu$m. The concave and planar dielectric mirrors have reflectivities of $>$99.99\% and 99.7\% respectively at the design wavelength of 637 nm, and reflection bands extending from 550 nm to 720 nm. The planar mirror is terminated with a low index layer to provide a field anti-node at its surface. The shortest cavity length achieved here has a mirror spacing of 1.11 $\mathbf{\mu}$m providing an additional 3 field intensity maxima of the TEM$_{00}$ mode between the mirror surfaces with $\lambda = 637$ nm, as shown in figure 1(d). We label this mode with the longitudinal index $q=4$, or the set of indices $\left(q,m,n\right)=\left(4,0,0\right)$. 

The NV$^-$ defects are located in nanodiamonds of average diameter 100 nm produced using high pressure high temperature methods, which are spin cast onto the planar mirror. Registration marks on the planar mirror created using a focused ion beam allow individual defects to be identified and characterised both in and out of the cavity. The two experimental geometries used in this report are displayed in figure 1. The `out-of-cavity' fluorescence is measured in the absence of a concave mirror and with the planar mirror as a substrate behind the nanodiamonds (fig 1a). For the `in-cavity' experiments the planar mirror is inverted and both excitation of the NV$^-$ centers (at $\lambda = 532$ nm) and collection of fluorescence is carried out through the planar mirror. The microcavity assembly comprises a set of piezoelectric actuators that provide full control of the cavity length and relative position of the planar and concave mirrors at cryogenic temperature (fig 1b). All measurements are carried out at 77K in a dry He exchange gas environment with the container supported in a liquid nitrogen bath cryostat \cite{Grazioso10}. Further details of the experimental apparatus can be found in section A of the supplementary information.

\section{Results and Discussion}

\begin{figure}[b]
	\begin{center}
		\includegraphics[width=14cm]{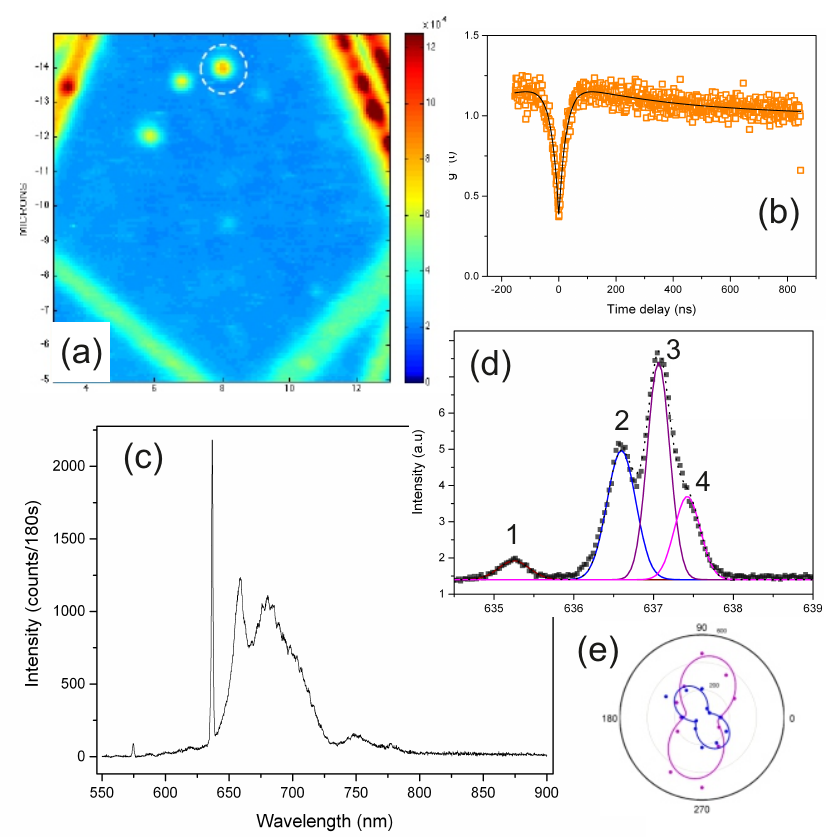} %depending on the latex compiler, you can omit the file extension
		\caption{Characterisation of the NV$^-$ center on the planar mirror. (a) Fluorescence image showing point-like emission from single NV$^-$ centers, and registration lines generated by focused ion beam Ga$^+$ implantation that allow re-location of the same NV defect in the two experimental configurations. (b) Photon correlation data from the NV$^-$ center measured, (c \& d) Fluorescence spectra from the NV$^-$ defect at $T=77$K, and (e) polarisation plots of peak 2 (blue) and peak 3 (purple) of the zero phonon line doublet.}
		\label{fig:example}
	\end{center}
\end{figure}

A low temperature fluorescence image recorded in the out-of-cavity configuration reveals well-dispersed nanodiamonds (fig 2a) with some instances of single centers recorded by the Hanbury Brown and Twiss method. The nanodiamond used in this study is circled, and its uncorrected photon intensity correlation data shown in fig 2b, revealing $g^{(2)}(0)=0.38$. Subtraction of background due to autofluorescence from the mirrors reduces this value to $g^{(2)}(0)=0.05$ implying that $\approx$ 97\% of the fluorescence is from a single emitter. Fluorescence spectroscopy of this NV$^-$ center reveals a ZPL spectrum that is dominated by a linearly polarised doublet (fitted peaks 2 and 3 in figure 2d) indicating a high level of strain that lifts the degeneracy of the orthogonal $^3E_x$ and $^3E_y$ excited state dipoles of the defect (fig 2c). The line widths of the individual ZPL components are approximately 0.4 nm, with Gaussian line shapes suggesting that they are dominated by inhomogeneous broadening due to spectral diffusion. Weak additional lines are seen, potentially due to an additional NV center in close proximity. The relative intensities of peaks 2 and 3 and the angle between their polarisations (fig 2e) indicate that the NV$^-$ defect axis is oriented at 49$^{\circ}$ relative to the optical axis of the cavity, and the two orthogonal transition dipoles for lines 2 and 3 are at angles of 39$^{\circ}$ and 24$^{\circ}$ to the plane of the mirror respectively (see supplementary information section B). The ZPL measured has DW = 0.044, as is typical for NV$^-$ centers. 

Figure 3 shows the measured fluorescence spectrum for the in-cavity configuration as a TEM$_{00}$ cavity mode is tuned through resonance with the ZPL. Clear enhancement of the ZPL emission is observed, providing a fully saturated photon count rate of 15 kc/s. This represents an experimental enhancement of the total ZPL signal by a factor of 2.5 compared with that recorded from the defect with the planar mirror alone (see supplementary information section C). 

\begin{figure}[t]
	\begin{center}
		\includegraphics[width=14cm]{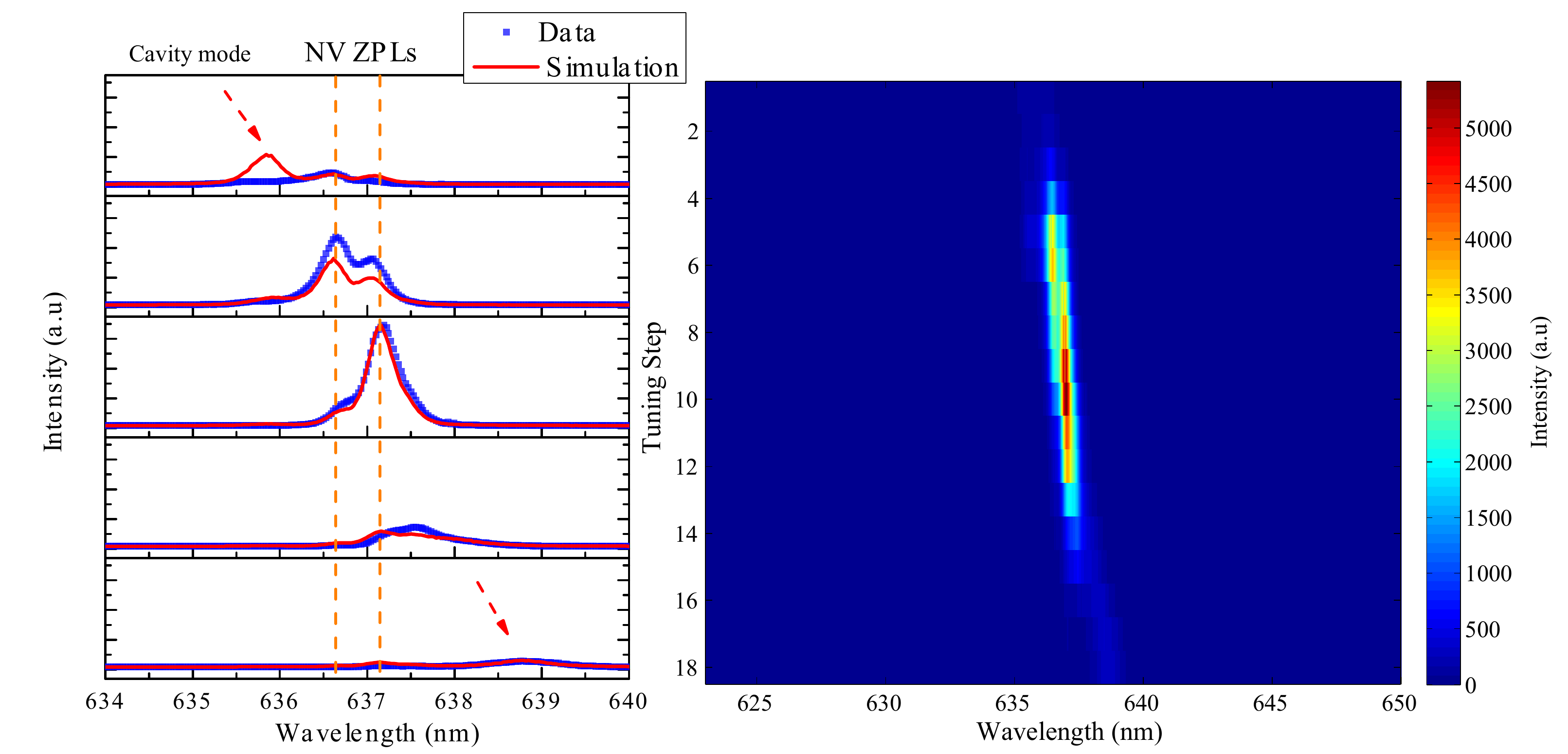} %depending on the latex compiler, you can omit the file extension
		\caption{In-situ tuning of a cavity mode through the ZPL resonance. (a) Selected spectra (blue) recorded at five different cavity positions, with a simulation of each spectrum calculated using equation 5 (red). (b) 2D false color-scale plot of the measured spectrum including all of the tuning steps recorded.}
		\label{fig:example}
	\end{center}
\end{figure}

Figure 4 shows a side-by-side comparison of the optical properties of the NV$^-$ center out of the cavity, with that observed in the cavity at optimal tuning to the ZPL. Figure 4a shows the spectra over the full range of NV- emission, illustrating the extent to which the coupled ZPL dominates the measured fluorescence, a result of the cavity having no other modes in this range that couple efficiently into the objective lens. Figure 4b shows a comparison between the fluorescence decay characteristics. The lifetime of the out-of-cavity defect is measured as ($30.8 \pm 0.6$) ns while that in the cavity is ($22.1 \pm 0.4$) ns, corresponding to a ($39.5 \pm 0.7$) \% increase in the emission rate. The slight deviation from a single exponential decay in the in-cavity data at longer delay time is suspected to be due to spectral instability of the cavity mode leading to inhomogeneous broadening. Figure 4c shows the photon autocorrelation data, which reveal a reduced $g^{(2)}(0)$ correlation value of 0.28 suggesting a slight improvement in the isolation of the single emitter when only the cavity-coupled ZPL is measured.

\begin{figure}[h]
	\begin{center}
		\includegraphics[width=16cm]{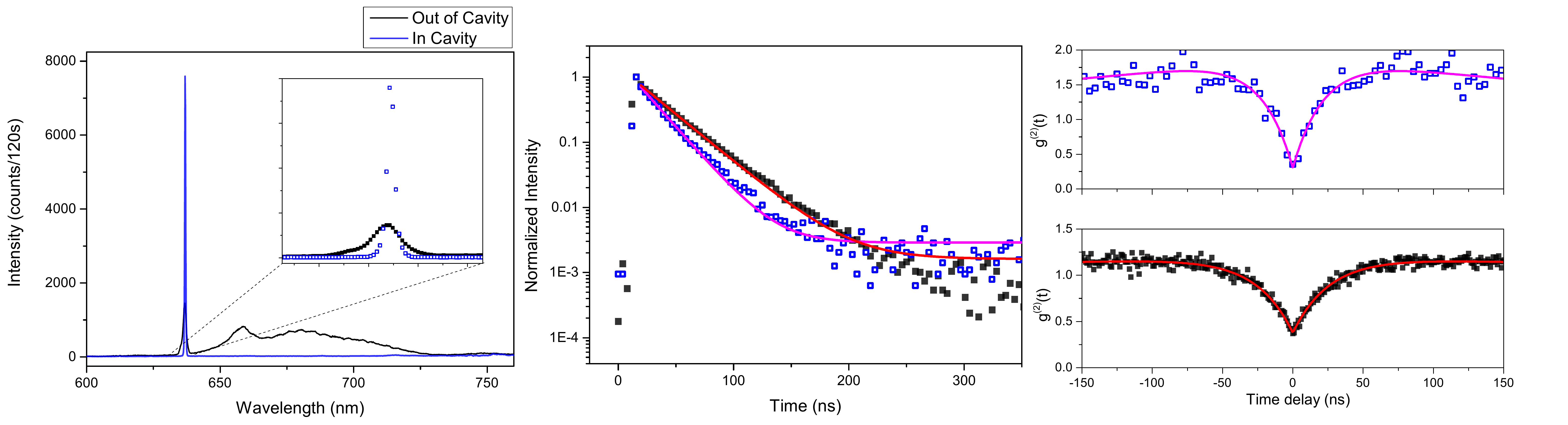} %depending on the latex compiler, you can omit the file extension
		\caption{Comparison of properties of the same NV$^-$ center in the in-cavity and out-of-cavity geometries. (a) fluorescence spectrum (b) fluorescence decay, and (c) photon correlation histogram. In (a) and (c) the excitation power is set to $P_{sat}$ as measured in intensity saturation measurements.}
		\label{fig:example}
	\end{center}
\end{figure}

We separate the analysis of our results into two parts - a semi-analytic treatment of the coupling of the ZPL to the TEM$_{00}$ mode and a numerical treatment of the coupling of the PSB to other cavity modes present. We begin with the ZPL coupling, for which we express the wavelength-dependent enhancement of the emission of a given dipole by a single cavity mode as:
\begin{equation}
F_{\mathbf{\mu}}(\lambda)= \xi_{\mathbf{\mu}} F^{max} \frac{ 1 }{1+4Q^2\left(\lambda/\lambda_{cav} -1\right)^2}
\end{equation}

where $F^{max}=\frac{3}{4\pi^2}\left(\frac{\lambda_{cav}}{n}\right)^3\frac{Q}{V_{mode}}$ is the maximum rate enhancement assuming perfect spatial alignment and orientation, $\xi_{\mathbf{\mu}}=\left(\frac{|\mathbf{\mu}.\mathbf{E}|}{|\mathbf{\mu}||\mathbf{E_{max}}|}\right)^2$ is the spatial overlap and orientation factor between the emitting dipole $\mathbf{\mu}$ and the cavity electric field $\mathbf{E}$, $\lambda_{cav}$ is the cavity wavelength and $Q$ is the cavity quality factor. As we are able to position the emitter at the electric field maximum we assume that $\mathbf{E}=\mathbf{E_{max}}$ so that $\xi_{\mathbf{\mu}}=\mathbf{cos}^2\left(\theta\right)$ where $\theta$ is the angle between the dipole and the plane of the mirror. 

The fractional increase in the total emission rate arising from the coupling with the ZPL is given by 
\begin{equation}
F_{ZPL}=\sum_{\mathbf{\mu}} n_{\mathbf{\mu}} \int\ d \lambda     S_{\mathbf{\mu}}(\lambda) F_{\mathbf{\mu}}(\lambda)
\end{equation}
where $n_{\mathbf{\mu}}$ are the branching factors of the two dipoles $\mathbf{\mu}$ in the excited state and $S_{\mathbf{\mu}}$ are their normalized free emission spectra collected over all directions. For peaks 2 and 3 of the ZPL doublet $n_{\mu}$ = 0.44 and 0.56 respectively (see supplementary information section B). $F_{ZPL}$ as defined includes the Debye Waller factor through the relative weight of $S_{\mu}$ corresponding to the ZPL. We rewrite equation 2 as:
\begin{equation}
F_{ZPL}  = F^{max} \left[\sum_{\mathbf{\mu}}   n_{\mathbf{\mu}} \xi_{\mathbf{\mu}} \right] \int\ d \lambda  \frac{ 1 }{1+4Q^2\left(\lambda/\lambda_{cav} -1\right)^2}  S_{axial}(\lambda) 
\end{equation}
where
\begin{equation}
S_{axial}(\lambda) = \frac{\sum_{\mathbf{\mu}}   n_{\mathbf{\mu}} \xi_{\mathbf{\mu}} S_{\mathbf{\mu}}(\lambda) }{\sum_{\mathbf{\mu}}   n_{\mathbf{\mu}} \xi_{\mathbf{\mu}}}
\end{equation}
is the normalized spectrum emitted along the cavity axis, to which the spectrum shown in figure 2(c) is a good approximation when appropriately scaled. The spectrum emitted into the cavity mode now reads
\begin{equation}
S_{cav}(\lambda) \propto \frac{ S_{axial}(\lambda)  }{1+4Q^2\left(\lambda/\lambda_{cav} -1\right)^2}  
\end{equation}

which, taking  $\delta\lambda_{cav}=0.7$ nm, reproduces well the measured tuning spectra in fig 3a. The same value of $\delta\lambda_{cav}$ used in equation 3 gives $F_{ZPL}=0.25$. This quantity is equal to the ratio of the overall ZPL emission rate into the cavity mode with the total emission rate in free space.

Based on the measured Debye Waller factor of 0.044 and the branching factor for peak 3 of 0.56, the cavity-induced enhancement of emission from peak 3 is found to be about a factor of 11. This compares with a value of 9.2 obtained using equation 1 with the effective Q factor $Q_{eff}=\lambda/\left(\delta\lambda_{cavity}+\delta\lambda_{emitter}\right)$, in which $\delta\lambda_{emitter}$ = 0.4 nm, the fitted line width for peak 3.

\begin{figure}[t]
	\begin{center}
		\includegraphics[width=10cm]{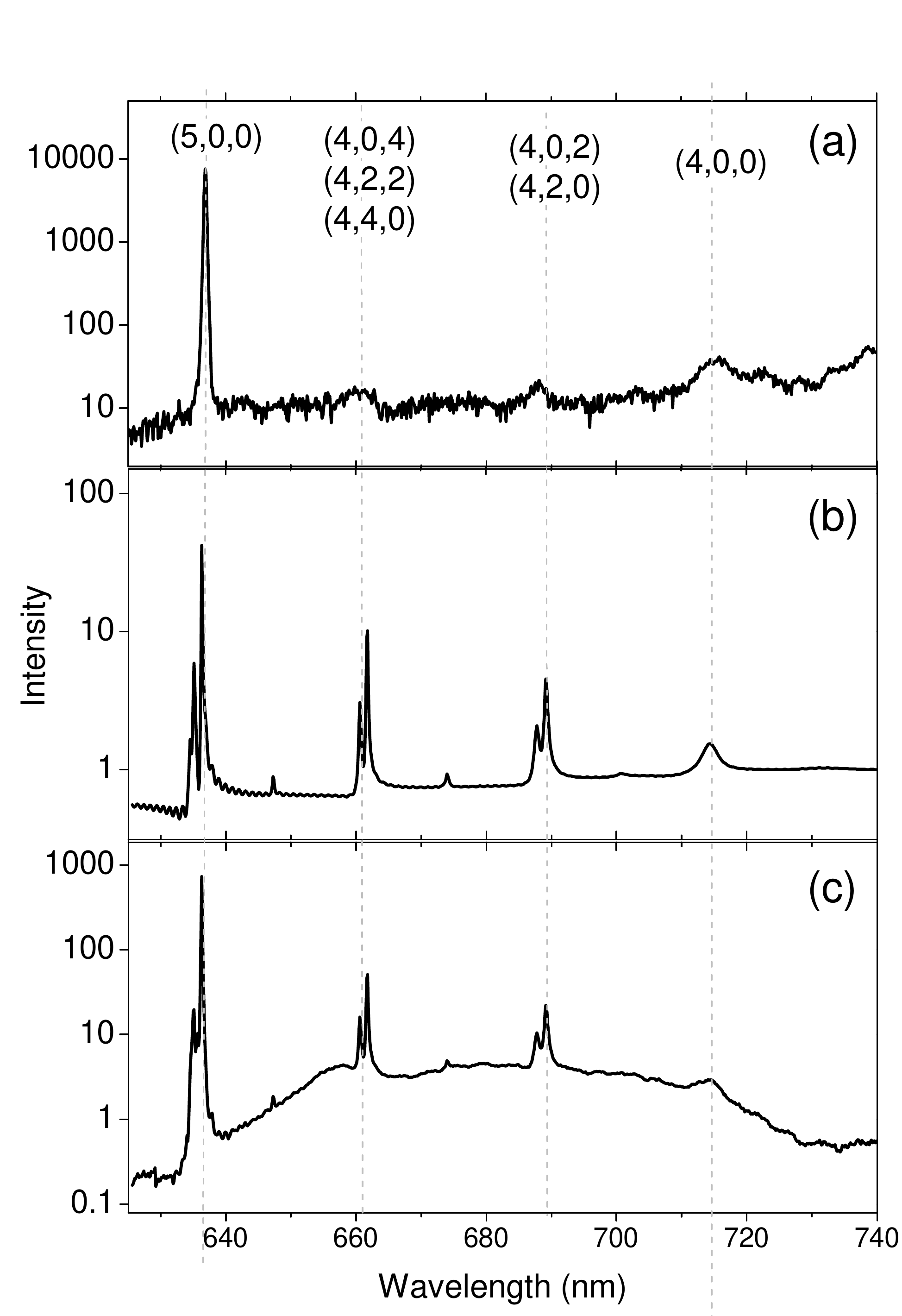} %depending on the latex compiler, you can omit the file extension
		\caption{Comparison between (a) the measured in-cavity spectrum with the (5,0,0) mode optimally coupled to the ZPL, (b) the FDTD simulation of the Purcell enhancement, and (c) the semi-empirical calculation of the emitted power density spectrum. Modes are identified by their $(q,m,n)$ for longitudinal index $q$ and transverse indices $m,n$.}
		\label{fig:example}
	\end{center}
\end{figure}

Modification of the optical density of states experienced by the phonon sideband was determined using numerical Finite Difference Time Domain (FDTD) calculations that reflect the full dielectric environment of the NV$^-$ center (see supplementary information section D for details). These calculations allow confirmation of the cavity parameters by matching the simulated mode spectrum to that measured, and a direct prediction of the change in emission rate that will occur between the in-cavity and out-of-cavity experimental configurations. Figure 5 shows semi-logarithmic plots of three cavity-coupled spectra with the (5,0,0) mode tuned to the ZPL. This measured spectrum (fig. 5a) is chosen as it reveals the positions of other modes coupling to the PSB and therefore provides a good reference by which to verify the geometrical parameters of the cavity. The relative Purcell factor (fig. 5b) and the resultant semi-empirical prediction for the emission power density spectrum of the NV$^-$ center in the cavity (fig 5c) are also shown. The PSB emission couples to cavity modes with longitudinal index $q=4$ and transverse indices 0, 2, and 4. The absence of observed coupling to modes with odd transverse indices suggests that the NV$^-$ center is well positioned on the cavity axis of symmetry where the electric field intensities of these modes drops to zero.

Integrating the emitted power density spectrum between 640 nm and 740 nm reveals the Purcell enhancement of the NV$^-$ center PSB emission to be $F_{P}^{PSB}=0.93$ (ie, a 7\% suppression of emission relative to the out-of-cavity geometry). The total change in emission rate for the NV$^-$ center is thus predicted to be $F_{ZPL}+F_{PSB} =1.14$ significantly lower than the measured value of 1.395. 

Figure 6a shows the calculated Purcell enhancement factors of the ZPL and PSB for different TEM$_{00}$ modes tuned into optimal resonance with the ZPL. The ZPL data points correspond to the enhancement of the emission rate corresponding to the entire ZPL: an estimate of the enhancement of peak 3 alone can be obtained by multiplying these values by $1/n_3=1.8$. Figure 6b reveals that the theory above underestimates the experimentally observed lifetime changes for each of these cavity lengths. The total emission rate is predicted to remain approximately the same for $q$ = 4, 5 and 6 because the increase in ZPL emission as the cavity length is reduced is compensated by a reduction in PSB emission.

\begin{figure}[t]
	\begin{center}
		\includegraphics[width=10cm]{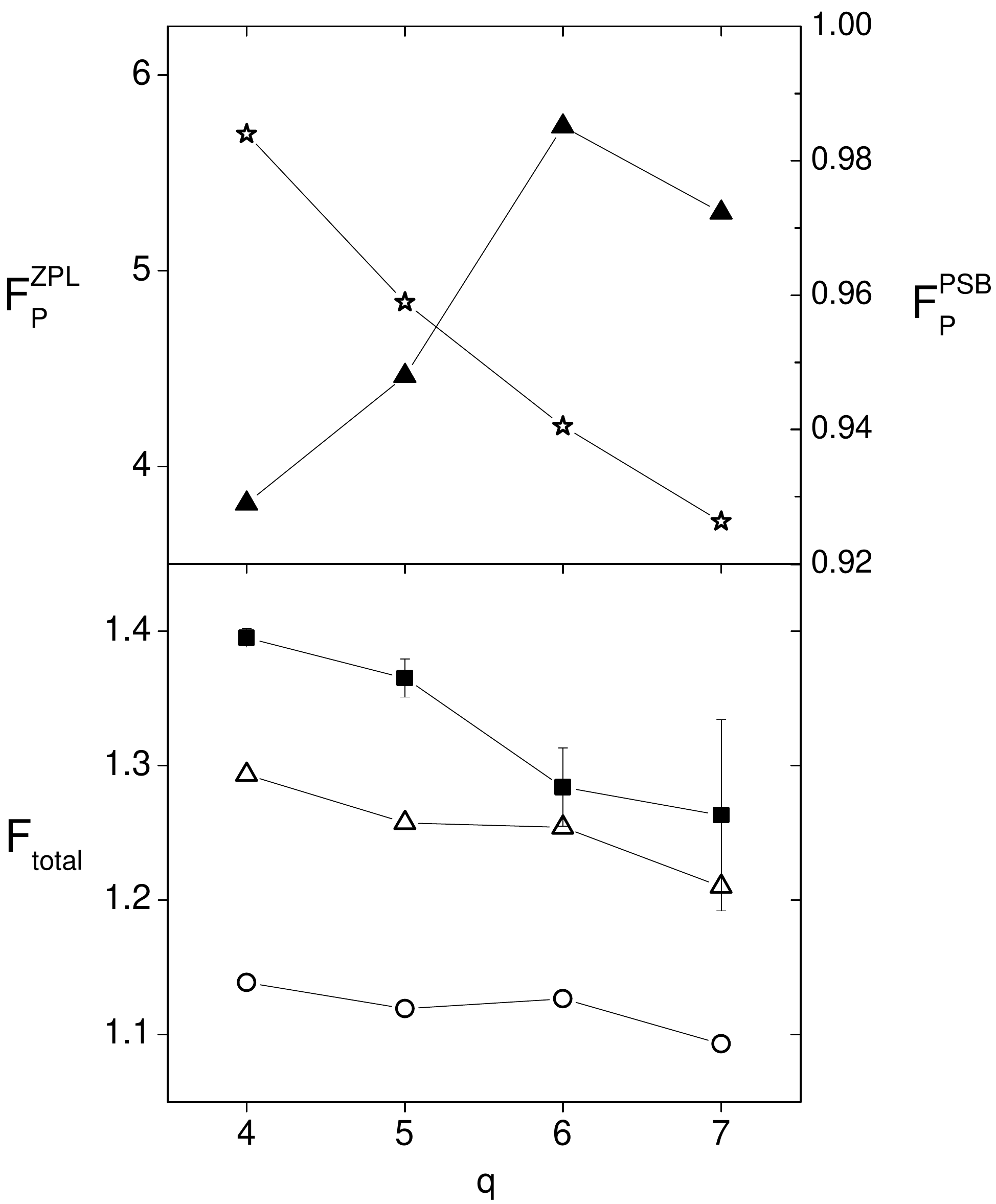} %depending on the latex compiler, you can omit the file extension
		\caption{Calculated and measured parameters as functions of the longitudinal mode index q. (a) Calculated Purcell enhancement factors for the zero phonon line (open stars) and the phonon sideband (closed triangles). (b) Comparison of the measured emission rate enhancement (solid squares) with the calculated values assuming a cavity line width of 0.7 nm and no inhomogeneous broadening (open circles) and with a cavity line width of 0.2 nm and inhomogeneous broadening of 0.5 nm (open triangles).}
		\label{fig:example}
	\end{center}
\end{figure}

We attribute the difference between the predicted and measured behavior to the presence of inhomogeneous broadening, both in the NV$^-$ center ZPL due to spectral drift and in the cavity linewidth due to mechanical instability in our apparatus. Under such conditions the fastest component of the measured emission decay curve corresponds to near-resonant alignment and will primarily reflect the homogeneous line widths. 
We introduce inhomogeneous broadening of the cavity mode into our analytic calculations above by convoluting the mode line width with a Gaussian function $g(\lambda_{cav})$. The change in the NV$^-$ emission rate due to the ZPL coupling is then given by (supplementary information section E):

\begin{equation}
F_{ZPL,inhom} =  \frac{\int d \lambda_{cav} ~ g(\lambda_{cav})  (1+F_{ZPL}(\lambda_{cav}) ) F_{ZPL}(\lambda_{cav})}{\int d \lambda_{cav} ~ g(\lambda_{cav})  F_{ZPL}(\lambda_{cav}) }
\end{equation}

A cavity line width of $\delta\lambda$ = 0.2 nm, corresponding to the value measured by transmission spectroscopy in a nominally identical cavity, combined with an inhomogeneous broadening of 0.5 nm, gives $F_{ZPL,inhom}$ = 0.364, whilst leaving the detuning spectra in figure 3  relatively unchanged. The overall changes in the NV emission rates are plotted in figure 6b, and are seen to agree more closely with, although still consistently underestimate, the measured values. 
We attribute the remaining discrepancy between the measured and modelled lifetimes to inhomogeneous broadening of the ZPL, which is more difficult to quantify and which the semi-empirical calculation method described above can not easily accommodate. A simple indication of the potential effect to the lifetimes in this experiment can be obtained from equation 1 however, in which a ZPL homogeneous line width of order 0.1 nm, consistent with values measured at this temperature in bulk materials \cite{Fu09}, and a cavity line width of 0.2 nm, give $F_{\mu}$ = 33.6 when resonantly tuned to peak 3. The resultant emission rate increase is $F_{tot}=1.71$, suggesting that the measured rate increase can indeed be accounted for by a combination of cavity and ZPL inhomogeneities. For completeness we note that an unknown parameter for our NV$^-$ center is its quantum efficiency (QE), which the above treatment assumes to be unity. Recent reports have demonstrated that nanodiamonds of the size used here can have QE as small as 0.3 \cite{Inam14}. The effect of reduced QE would be to require higher still Purcell factors to achieve the measured lifetime change, since the enhancement acts only on the radiative term.

In conclusion we have shown the controlled coupling of the ZPL of a single NV$^-$ center in nanodiamond to an open cavity at cryogenic temperatures. This cavity system shows significant potential for interfacing the NV$^-$ center with photonic networks and performing quantum operations and measurements. The degree of enhancement is currently limited by the line width of the NV$^-$ centers in the nanodiamond, and by the cavity line width as determined by scattering and instability of the low temperature cavity assembly. Improvements in these areas are readily achievable and are expected to produce much larger enhancements than those reported here. Open cavity quality factors exceeding $10^6$ have been demonstrated \cite{Hunger10}, and modest further reductions in cavity mode volume are clearly also possible. Single NV$^-$ defects implanted into bulk diamond at depths of 100 nm can offer ZPL line widths as narrow as 27 MHz, \cite{Chu14} and electron spin coherence times $>100 \mu$s. Resonant ZPL coupling of NV$^-$ centers situated in diamond membranes to open cavities can thus result in enhancement of the emission rate into the ZPL by a factor of $>10^3$, leading to an effective Debye Waller factors approaching unity and indistinguishable photons with lifetimes of a few hundred picoseconds. Such projections suggest that the experimental configuration demonstrated here is an attractive route towards an efficient spin/photon interface and to the construction of scalable quantum processors.

This work was funded by the European Union Seventh Framework Programme (FP7/2007-2013) under grant agreement no 61807. SJ acknowledges support from the United Kingdom Engineering and Physical Sciences Research Council.

\section*{Supplementary Information}

\subsection*{A) Experimental Methods}

\begin{figure}[b]
	\begin{center}
		\includegraphics[width=14cm]{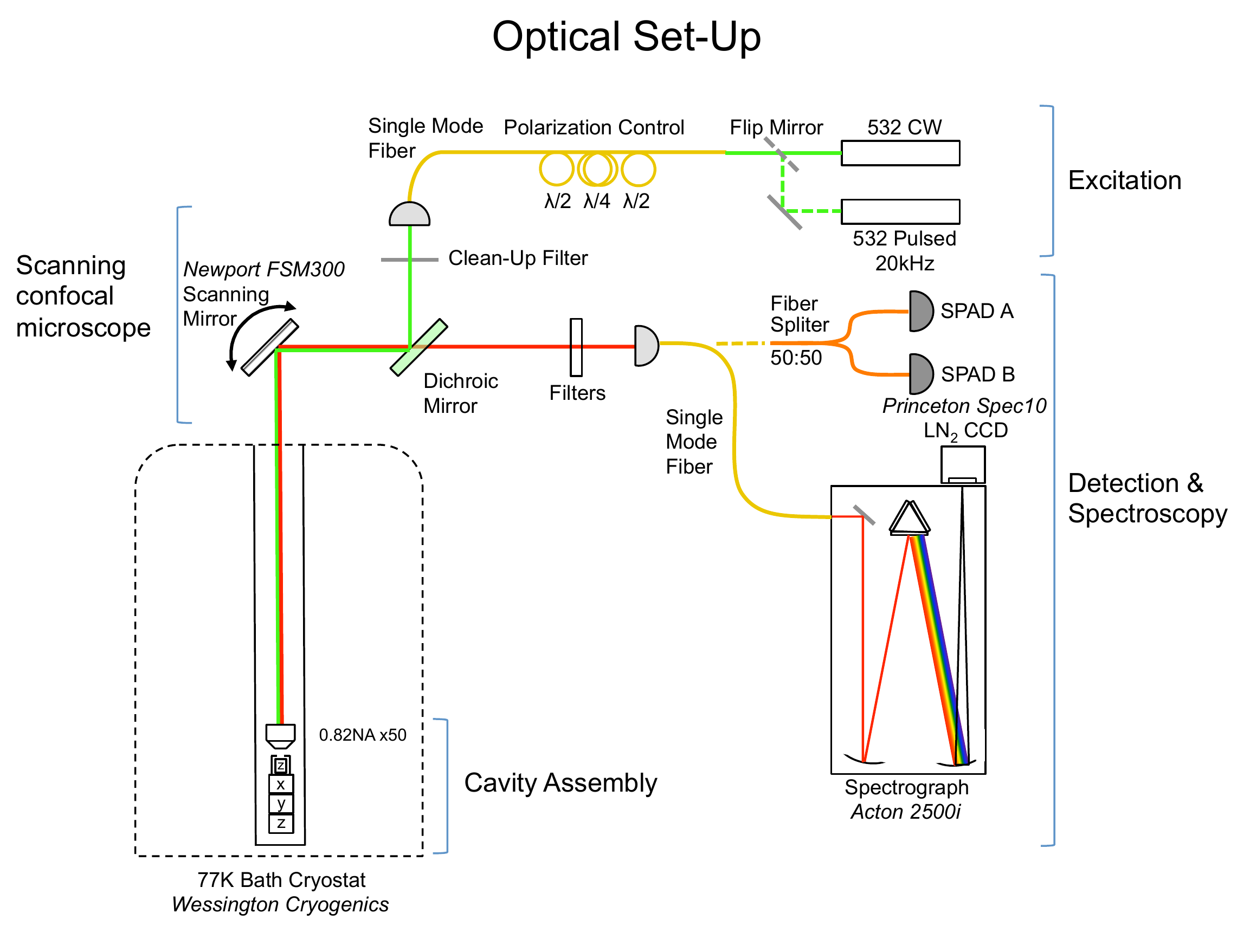} %depending on the latex compiler, you can omit the file extension
		\caption{Optical set-up for low temperature cavity experiments}
		\label{fig:optics}
	\end{center}
\end{figure}

Cavities are milled into fused silica substrates (UQG Optics) using a focused ion beam (FEI FIB200). High-reflectivity mirror stacks are then deposited on the substrates. The planar mirror is coated in-house using alternating $\lambda/4n$ layers of SiO$_2$/TiO$_2$ to achieve 99.7\% reflectivity at $\lambda$ = 637nm. The planar mirror is terminated with the low index materials such that there is an anti-node of the electric field at the surface for emitter-cavity coupling. The concave mirror was coated at LaserOptik GmbH, with 20 pairs of SiO$_2$/Ta$_2$O$_5$ to provide R>99.99\%, such that optical coupling is preferentially through the planar mirror. Nanodiamond solutions (100nm - 0.1mg/ml) are spin coated onto the planar mirror substrates prior to cavity coupling. Optical characterisation is carried out with a home-built beam-scanning confocal microscope employing a fast-steering mirror (Newport FSM300), outlined in figure \ref{fig:optics}. 532nm CW excitation is used for imaging and spectroscopy, with 532nm pulsed excitation available for PL lifetime experiments (Teem Photonics-SNG-20F-1SO, repetition rate = 20KHz). Excitation is coupled through a single mode fiber (Thorlabs SM460HP) with polarisation control (Thorlabs FPC030). The cavity apparatus is situated in a dry He-exchange gas environment (50 mbar), immersed in a liquid nitrogen bath cryostat with optical access. The apparatus consists of a custom sample stage, allowing independent piezo actuation of the mirror substrates. The planar mirror substrate has all the translational degrees of freedom (Attocube: 2 x ANPx100, 1 x ANPz100), whilst the cavity substrate can only move vertically (Attocube ANPz30). All stepper motors are driven with an Attocube ANC300 control module. A low temperature compatible achromatic objective (Attocube ASWDO x50 0.82NA) is used for optical excitation and collection. Spectral filters can be placed in the collection path. Cavity coupled HBT, Lifetime and power saturation measurements are taken in the 633-647nm spectral window using an additional band pass filter (Semrock FF01-640/14-25). 
Fluorescence is coupled into single photon avalanche detectors (Perkin-Elmer SPCM-AQRH) via a single mode fiber (Thorlabs SM600). A 1x2 50:50 fiber splitter (Thorlabs FCMM50-50A) with second single photon detector are used for HBT measurements. An Edinburgh Instruments TCC900 card provides the correlation electronics for the HBT and lifetime measurements. The output fiber can also be coupled to a spectrograph (Acton 2500i), equipped with a liquid nitrogen cooled CCD (Princeton Spec10:100B). 

\subsection*{B) Calculation of the ZPL dipole orientations}
The orientations of the dipoles for the two strain-split ZPL transitions are calculated by determining the polar angle of the NV axis relative to the optical axis ($\theta$) and the rotation of the dipoles about the NV axis ($\beta$) which will lead to the projections on the observation plane. 

A thermal distribution between the populations of the two excited states comprising the ZPL doublet is also assumed, consistent with the findings of Fu et al \cite{Fu09}. The energy splitting of peaks 2 and 3 is 1.5 meV, so that the population ratio for the levels responsible for peaks 2 and 3 at 77K is 0.8:1. The measured intensity ratio between peaks 2 and 3 due to the projection of this thermally distributed dipole pair onto the measurement plane is 0.58:1, so the equivalent projection of a circle formed by two perpendicular dipoles of equal strength would result in a ratio of $R=0.73$. 

The polar angle $\theta$ is obtained from the sum of the polar intensity distributions of peaks 2 and 3, whereby the extrema have the following dependence on $\theta$. 
  
\begin{equation}
I_{min} = I_{max}cos^2(\theta)
\end{equation}
  
The combined rotation matrix for the axial, then polar, rotation is 

\begin{equation}
A = \left( \begin{array}{ccc}
cos(\beta) & -sin(\beta) & 0\\
cos(\theta)sin(\beta) & cos(\theta)cos(\beta) & -sin(\theta)\\
sin(\theta)sin(\beta) & sin(\theta)cos(\beta) & cos(\theta)
\end{array} \right)
\end{equation}

Applying this rotation to unit vectors X \& Y, and projecting the resultant vectors onto the measurement plane, gives
\[
X^{'2}=1-sin^2(\theta)sin^2(\beta)\\ \qquad
Y^{'2}=1-sin^2(\theta)cos^2(\beta)
\]

Where the square of the dipole vector have been taken to obtain the intensity. Knowing $\theta$ and the ratio R between these intensities, allows $\beta$ to be determined. Finally, the angles $\phi_{X'}$ \& $\phi_{Y'}$ between the rotated dipole vectors and the observation plane are found by taking the dot product of the rotated dipole vectors and their projections, leading to 

\[
\phi_{X'} = arccos \left( \sqrt{1-sin^2(\theta)sin^2(\beta)}\right) \\ \qquad
\phi_{Y'} = arccos \left( \sqrt{1-sin^2(\theta)cos^2(\beta)}\right)
\]

\subsection*{C) Comparison of excitation conditions for the two experimental geometries}

For comparison of the emission intensities in the out-of-cavity and in-cavity geometries it is necessary to establish equivalent excitation conditions in the two cases. To do this we measured the dependence of the emission intensity $I$ on excitation power $P$ to record the emission saturation curve of the color center in each case. These data are shown in figure 8. We then fitted these curves to the saturation function

\[
I = \frac{I_{sat}\ P}{P_{sat}+P}
\]

The fitting parameters $I_{sat}$ and $P_{sat}$ are the fully saturated photon count rate and the characteristic saturation power for excitation respectively. In the out-of-cavity geometry we find that $I_{sat}=$ 154 kc/s for the whole NV centre (6.78 kc/s for the ZPL only) and $P_{sat}=1.02$ mW, whilst that in the $q=4$ cavity (data in figures 3 and 4) yields $I_{sat}=$ 15.1 kc/s for the ZPL only and $P_{sat}=1.89$ mW. The spectra shown in figure 4 use an excitation power of $P_{sat}$ in each case.

\begin{figure}[h]
	\begin{center}
		\includegraphics[width=14cm]{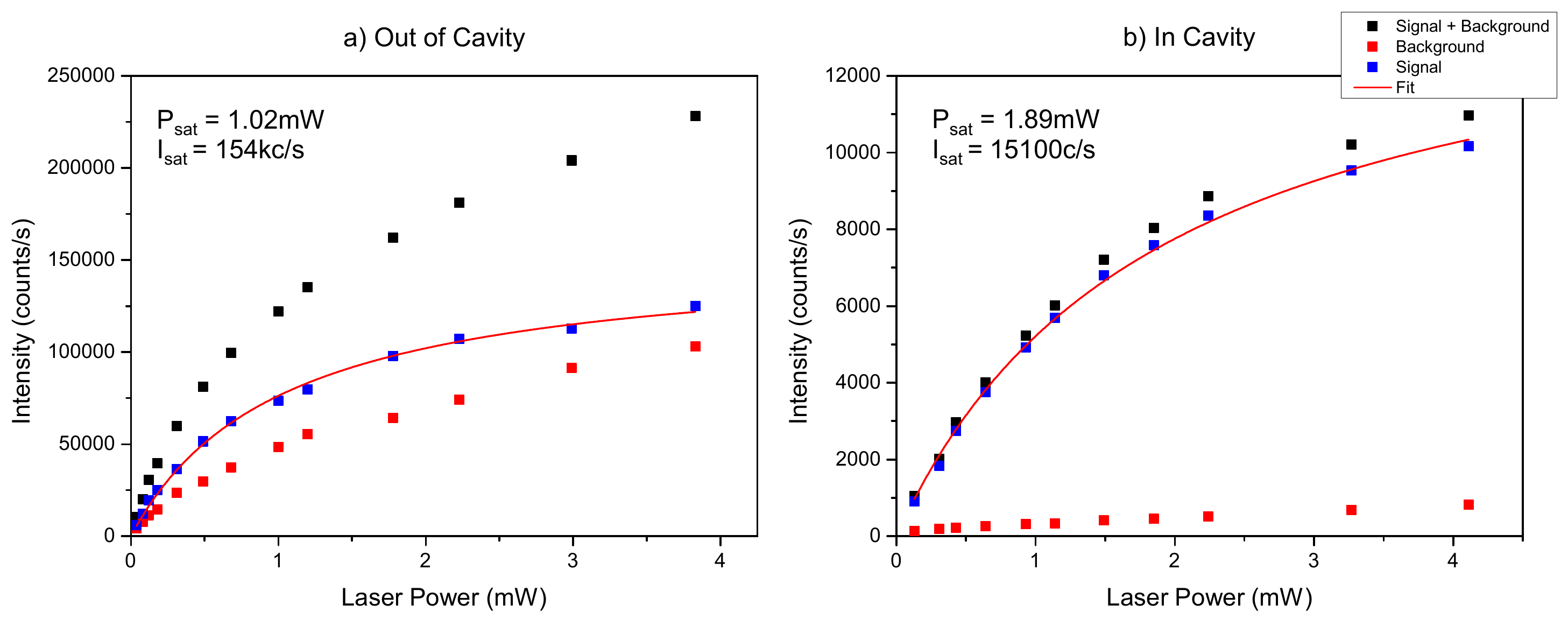} %depending on the latex compiler, you can omit the file extension
		\caption{Power Saturation measurement into the cavity coupled ZPL with mode number q=4.}
		\label{fig:Psat}
	\end{center}
\end{figure}

\subsection*{D) Finite Difference Time Domain simulations}

We used Lumerical Solutions FDTD software to perform numerical simulations of a dipole emitting into the cavity. Firstly, by reproducing the measured mode structure using these simulations we confirm the cavity geometry. The concave mirror is modelled as vertically stacked films of $\lambda/4$ vertical height, based on cross-sectional SEM measurements such as the one shown in figure 1(a). Refractive indices of the SiO$_2$ and Ta$_2$O$_5$ layers are 1.52 and 2.10 respectively. The concave mirror is situated as having 15 Bragg pairs rather than the 20 in the real device, as this allows faster calculations with negligible affect on the results. The dipolar source is positioned on the axis of symmetry, 20 nm from the planar mirror surface, and within a dielectric sphere of diameter 100 nm and refractive index 2.4 in contact with the planar mirror to simulate the nanodiamond. Separate calculations are performed with dipoles at angles of 18$^{\circ}$ and 39$^{\circ}$ to the planar mirror representing peaks 3 and 2 respectively. We confirmed that the results of the calculation are insensitive to the position of the source within the nanodiamond and on the exact size of the nanodiamond. The finite element Yee cells have a minimum size of 5 nm to accommodate the nanodiamond and the contours of the curved mirror. The electromagnetic field was allowed to propagate and decay for 10 ps in 5 fs increments, corresponding to a simulation time of about 12 hours. 

Mode volumes were calculated both using the simple analytic formula for a Gaussian beam and numerical integration of the simulated FDTD field intensity. The analytic expression used is $V_{\mathrm{mode}}=\frac{\lambda z_R L}{4}$ where $z_R = L \sqrt{\left(\frac{\beta}{L}-1\right)}$ is the Rayleigh range, $L$ is the optical length of the cavity and $\beta$ is the radius of curvature of the concave mirror. The integration used is

\[
V_{\mathrm{mode}} =\frac{\int Re[\epsilon]\left|E\right|^2 \, \mathrm{d}V }{\left(Re[\epsilon]\left|E\right|^2\right)_{dipole}}
\]

\subsection*{E) Calculation of inhomogeneous broadening}

For an inhomogeneously broadened cavity mode, the total intensity is given by an average over the different cavity positions
\begin{equation}
I(t) = \int d \lambda_{cav} ~ g(\lambda_{cav})  A(\lambda_{cav} ) e^{-\gamma(\lambda_{cav})t }
\end{equation} 
where $g(\lambda_{cav})$ is the inhomogeneous distribution of the cavity position, taken as a normalized Gaussian.
For a given cavity position, the fraction of light emitted in the cavity mode is proportional to the measured intensity integrated over a full decay event
\begin{equation}
\beta(\lambda_{cav} ) = \frac{F_{ZPL}(\lambda_{cav})}{1+F_{ZPL}(\lambda_{cav})} \propto A(\lambda_{cav}) /\gamma(\lambda_{cav})
\end{equation}

The decay rate is then found by differentiating equation 9: 

\begin{equation}
\gamma_{inhom} = - \frac{dI}{I dt}\biggr\vert_{t=0} = \frac{\int d \lambda_{cav} ~ g(\lambda_{cav})  \gamma^2(\lambda_{cav} ) \beta(\lambda_{cav} )}{\int d \lambda_{cav} ~ g(\lambda_{cav})  \gamma(\lambda_{cav} ) \beta(\lambda_{cav} )}
\end{equation} 

so that

\begin{equation}
F_{inhom} = \frac{\gamma_{inhom}}{\gamma_0} = \frac{\int d \lambda_{cav} ~ g(\lambda_{cav})  (1+F_{ZPL}(\lambda_{cav}) )^2 \beta(\lambda_{cav} )}{\int d \lambda_{cav} ~ g(\lambda_{cav})  (1+F_{ZPL}(\lambda_{cav}) ) \beta(\lambda_{cav} )}
\end{equation} 

whereby substituting back in for $\beta(\lambda_{cav})$ from equation 10 gives

\begin{equation}
F_{inhom} =  \frac{\int d \lambda_{cav} ~ g(\lambda_{cav})  (1+F_{ZPL}(\lambda_{cav}) ) F_{ZPL}(\lambda_{cav})}{\int d \lambda_{cav} ~ g(\lambda_{cav})  F_{ZPL}(\lambda_{cav}) }
\end{equation}



\vspace{20pt}

\begin{thebibliography}{}

\bibitem{Dutt07} M. V. Gurudev Dutt, L. Childress, L. Jiang, E. Togan, J. Maze, F. Jelezko, A. S. Zibrov,
P. R. Hemmer and M. D. Lukin, Science, \textbf{2007}, 316. 1312.
\bibitem{Childress08} L. Childress, M. V. Gurudev Dutt, J. M. Taylor, A. S. Zibrov, F. Jelezko, J. Wrachtrup, P. R. Hemmer and M. D. Lukin, Science \textbf{2006}, 314, 281.
\bibitem{Balasub08} G. Balasubramanian, I. Y. Chan, R. Kolesov, M. Al-Hmoud, J. Tisler, C. Shin, C. Kim, A. Wojcik, P. R. Hemmer, A. Krueger, T. Hanke, A. Leitenstorfer, R. Bratschitsch, F. Jelezko and J. Wrachtrup, Nature \textbf{2008}, 455, 648.
\bibitem{Rondin14} L. Rondin, J-P Tetienne, T. Hingant, J.-F. Roch, P. Maletinsky and V. Jacques, Rep. Prog. Phys. \textbf{2014}, 77, 056503.
\bibitem{Toyli13} D. M. Toyli, C. F. de las Casas, D. J. Christle, V. V. Dobrovitski and D. D. Awschalom, Proc. Nat. Assoc. Sci. \textbf{2013}, 110, 8419.
\bibitem{Waldherr14} G. Waldherr, Y. Wang, S. Zaiser, M. Jamali, T. Schulte-Herbr\"{u}ggen, H. Abe, T. Ohshima, J. Isoya, J. F. Du, P. Neumann and J. Wrachtrup, Nature \textbf{2014}, 506, 204.
\bibitem{Barrett05} S. Barrett and P. Kok, Phys. Rev. A. \textbf{2005}, 71, 060310(R).
\bibitem{Bernien13} H. Bernien, B. Hensen, W. Pfaff, G. Koolstra, M. S. Blok, L. Robledo, T. H. Taminiau, M. Markham, D. J. Twitchen, L. Childress and R. Hanson,  Nature \textbf{2013}, 497, 7447, 86.
\bibitem{Raussendorf01} R. Raussendorf and H. J. Briegel, Phys. Rev. Lett. \textbf{2001}, 86, 5188.
\bibitem{Benjamin09} S. C. Benjamin, B. W. Lovett and J. M. Smith, Laser Photon. Rev. \textbf{2009}, 3, 556-574.
\bibitem{Faraon11} A. Faraon, P. E. Barclay, C. Santori, K.-M. C. Fu and R. G. Beausoleil, Nature Photon. \textbf{2011}, 5, 301.
\bibitem{Faraon12} A. Faraon, C. Santori, Z. Huang, V. M. Acosta and R. G. Beausoleil, Phys. Rev. Lett. \textbf{2012}, 109, 033604.
\bibitem{Li15} L. Li, T. Schr\"{o}der,, E. H. Chen, M. Walsh, I. Bayn, J. Goldstein, O. Gaathon, M. E. Trusheim, M. Lu, J. Mower, M. Cotlet, M. L. Markham, D. J. Twitchen and D. Englund, Nature Commun. \textbf{2015}, 6, 6173.
\bibitem{Trupke05} M. Trupke, E. A. Hinds, S. Eriksson, E. A. Curtis, Z. Moktadir, E. Kukharenka and M. Kraft, Appl. Phys. Lett. \textbf{2005}, 87, 211106.
\bibitem{Steinmetz06} T. Steinmetz, Y. Colombe, D. Hunger, T. W. H\"{a}nsch, A. Balocchi, R. J. Warburton and J. Reichel, Appl. Phys. Lett. \textbf{2006}, 89, 111110.
\bibitem{Dolan10} P. R. Dolan, G. M. Hughes, F. Grazioso, B. R. Patton, and J. M. Smith, `Femtoliter tunable optical cavity arrays', Optics Lett. \textbf{2010}, 35, 3556.
\bibitem{Hunger10} D. Hunger, T. Steinmetz, Y. Colombe, C. Deutsch, T. W. H\"{a}nsch, and J. Reichel, New J. Phys \textbf{2010}, 12, 065038.
\bibitem{Albrecht13} R. Albrecht, A. Bommer, C. Deutsch, J. Reichel and C. Becher, Phys. Rev. Lett. \textbf{2013}, 110, 243602.
\bibitem{Albrecht14} R. Albrecht, A. Bommer, C. Pauly, F. M\"{u}cklich, A. W. Schell, P. Engel, T. Schr\"{o}der, O. Benson, J. Reichel and C. Becher, Appl. Phys. Lett \textbf{2014}, 105, 073113.
\bibitem{Kaupp13} H. Kaupp, C. Deutsch, H. C. Chang, J. Reichel, T. W. H\"{a}nsch and D. Hunger, Phys. Rev. A \textbf{2013}, 88, 053812.
\bibitem{Grazioso10} F. Grazioso, B. R. Patton and J. M. Smith, Rev. Sci. Inst. \textbf{2010}, 81, 093705.
\bibitem{Di12} Z. Y. Di, H. V. Jones, P. R. Dolan, S. M. Fairclough, M. B. Wincott, J. Fill, G. M. Hughes and J. M. Smith, New J. Phys. \textbf{2012} 14, 103048.
\bibitem{RiedrichMoller14} J. Riedrich-M\"{o}ller, C. Arend, C. Pauly, F. M\"{u}cklich, M. Fischer, S. Gsell, M. Schreck and C. Becher, Nano Lett. \textbf{2014}, 14, 5281.
\bibitem{Fu09} K.-M. C. Fu, C. Santori, P. E. Barclay, L. J. Rogers, N. B. Manson and R. G. Beausoleil, Phys. Rev. Lett. \textbf{2009}, 103, 256404.
\bibitem{Inam14} F. A. Inam, M. J. Steel and S. Castelletto, Diamond and Related Materials \textbf{2014}, 45, 64-69.
\bibitem{Chu14} Y. Chu, N.P. de Leon, B. J. Shields, B. Hausmann, R. Evans, E. Togan, M. J. Burek, M. Markham, A. Stacey, A. S. Zibrov, A. Yacoby, D. J. Twitchen, M. Loncar, H. Park, P. Maletinsky and M. D. Lukin, Nano Lett. \textbf{2014}, 14, 1982-1986. 


\end{thebibliography}
\end{document}